\documentclass[aps,prl,twocolumn,showpacs,amsmath,amssymb]{revtex4}
\usepackage{epsfig}
\usepackage{bm}
\usepackage{color}

\begin{document}

\title{Quantum state tomography with quantum shotnoise}
\author{P. Samuelsson$^1$ and M. B\"uttiker$^2$}
\affiliation{$^1$Division of Solid State Theory, Lund University,
S\"olvegatan 14 A, S-223 62 Lund, Sweden \\ $^2$D\'epartement de
Physique Th\'eorique, Universit\'e de Gen\`eve, CH-1211 Gen\`eve 4,
Switzerland}

\date{\today}

\begin{abstract}
We propose a scheme for a complete reconstruction of one- and
two-particle orbital quantum states in mesoscopic conductors. The
conductor in the transport state continuously emits orbital quantum
states. The orbital states are manipulated by electronic beamsplitters
and detected by measurements of average currents and zero frequency
current shotnoise correlators. We show how, by a suitable complete set
of measurements, the elements of the density matrices of the one- and
two-particle states can be directly expressed in terms of the currents
and current correlators.

\end{abstract}

\pacs{03.67.Mn,42.50.Lc,73.23.-b}
\maketitle 

According to the standard interpretation of quantum mechanics the
wavefunction, or more generally the density matrix, determines the
probabilities for the possible outcomes of any measurement on the
quantum state.  To completely characterize the wavefunction of the
state is therefore of fundamental interest \cite{Pauli}. It is however
impossible to infer anything about an unknown state from a single
measurement, a complete characterization requires an ensemble of
identically prepared states and the measurement of a complete set of
observables on the state \cite{Fano}. A reconstruction of the quantum
state wavefunction via such a series of measurements is known as
Quantum State Tomography (QST) \cite{QSTrevs}.

Initially, QST was performed experimentally on the discrete angular
momentum state of an electron in an hydrogen atom
\cite{Ashburn}. During the last decade QST has been performed on e.g.
the quantum state of squeezed light \cite{Smithey}, the vibrational
state of a molecule \cite{Dunn}, the motional state of trapped ions
\cite{Liebfried} and of atomic wavepackets \cite{Kurtsiefer}. Recently
there has been an interest in QST of two-particle states in the
context of quantum information processing. The entanglement of a
quantum state, a potential resource for quantum information
processing, is characterized by the density matrix of the state. The
quantum state of polarization entangled pairs of photons has been
reconstructed using QST \cite{EntQST}.
 
To date, no QST has been performed on quantum states in solid state
systems.  Very recently a theoretical scheme \cite{Nori} was developed
for solid state two-levels systems, qubits, appropriate for e.g. the
macroscopic superposition state in superconducting qubits and the
spin-state of electrons in quantum dots. The set of measurements
necessary to reconstruct the state involves controlled rotations and
detection of the individual qubits. For coupled qubits, where
entanglement between the qubits is of interest, such measurements are
highly involved and have not been demonstrated.
  
In this paper we take a different approach and present a scheme for
QST of discrete single and two-particle orbital quantum states in
mesoscopic conductors. The orbital quantum states \cite{Orb,QH1,QH2}
are continuously emitted from the conductor during transport, making a
long time measurement equivalent to an average over an ensemble of
states. The orbital states can be manipulated by electronic
beamsplitters, experimentally available \cite{BS1,BS2}, and detected
by measurements of average currents and zero frequency current
correlators, shotnoise \cite{Buttnoise,Buttrev}. This scheme, with all
components experimentally realizable in e.g. Quantum Hall systems
\cite{BS2,QHexp}, allows for a complete characterization of the
quasiparticle quantum state in mesoscopic conductors.

The key question for any QST is: what quantum states with interesting
properties can be investigated with accessible experimental technics?
In mesoscopic conductors, one typically measures electrical currents
and current correlators. In several recent works
\cite{Sukh,Cht,Orb,QH1,QH2}, it has been shown theoretically that
quantum correlations, entanglement, between two spatially separated
particles can be investigated via current correlation measurements. In
particular, in Refs. \cite{Orb,QH2} it was shown how entangled orbital
quasiparticle states could be generated, manipulated with
experimentally available electronic \cite{BS1,BS2} beamsplitters and
detected via current correlation measurements. Orbital one- and
two-particle states investigated via current and current correlations
are thus natural candidates for QST in mesoscopic conductors.
\begin{figure}[b]   
\centerline{\psfig{figure=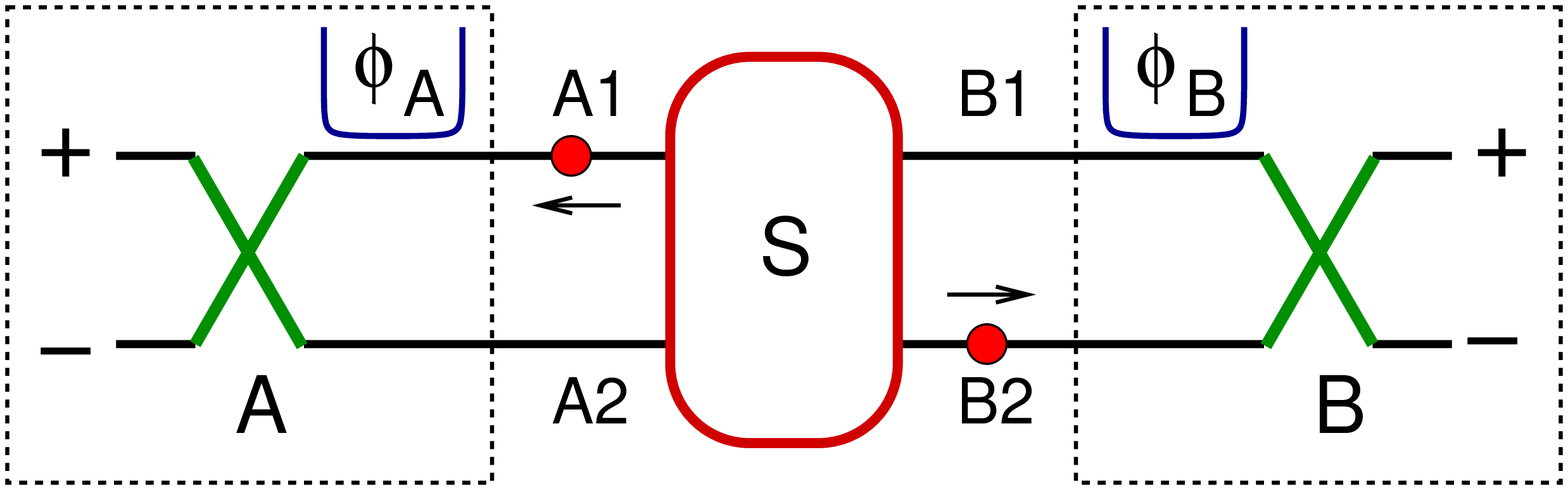,width=8.5cm}}
\caption{Schematic of the setup. A mesoscopic conductor acting as a
source S (red box) is connected via four leads, $A1,A2,B1$ and
$B2$, to regions A and B (dashed boxes), each containing two
reservoirs, $+$ and $-$, a beamsplitter (green cross) and a sidegate
(blue open box) which induces a phaseshift $\phi_A$ or $\phi_B$. Two particles (red dots) emitted from the source are shown.}
\label{figcond}
\end{figure}

A generic setup for such orbital QST is shown in Fig. \ref{figcond}. A
mesoscopic conductor $S$ acts as a source for orbital quantum
states. The source is connected via four single mode leads $A1,A2,B1$
and $B2$ to two regions, $A$ and $B$, where the emitted state is
manipulated and detected. The mesoscopic source has one or more
reservoirs biased at $eV$ and an arbitrary number of reservoirs kept
at ground. We note that two-particles effects are only present for two
or more reservoirs biased \cite{Buttnoise}. The temperature is taken
to be zero. It is assumed that the scattering in the conductor is
elastic, however arbitrary debasing inside the conductor can be
accounted for.

The regions $A$ and $B$ each contain an electronic beamsplitter
\cite{BS1,BS2} and an electrostatic sidegate (see e.g. \cite{QHexp})
to induce a phaseshift, $\phi_A$ or $\phi_B$, by modifying the length
of the lead. The beamsplitters, taken to be reflectionless, are
further connected to two grounded reservoirs $+$ and $-$ where the
current is measured. The combined beamsplitter-sidegate structure can
be characterized by a scattering matrix, for e.g. $A$ given by
\begin{eqnarray}
S_A=\left(\begin{array}{cc} \sqrt{R_A}e^{i\varphi_{A2}} &
\sqrt{T_A}e^{i(\varphi_{A3}-\phi_{A})} \\ \sqrt{T_A}e^{i(\varphi_{A1}+\varphi_{A2})} &
-\sqrt{R_A}e^{i(\varphi_{A1}+\varphi_{A3}-\phi_A)\ } \end{array}\right).
\label{smat}
\end{eqnarray}
The transmission probability $T_A=1-R_A$ can be controlled via
electrostatic gating \cite{BS1,BS2,QHexp}. The phases
$\varphi_{Ai},~i=1,2,3$ picked up when scattering at the beamsplitter
are however assumed to be uncontrollable but fixed during the
measurement.

The quantum state emitted by the mesoscopic source is in the general
case a manybody state, it is a linear superposition of states with
different number of particles \cite{SSQI}. However, one- and
two-particle observables such as current and noise are only sensitive
to the one- and two-particle properties of the state. These properties
are quantified by the reduced density matrix, which thus is the object
of interest. Only in some special cases, typically in conductors in
the tunneling limit \cite{Orb,QH1,QH2}, are the emitted states true
one or two-particle states. In the presence of dephasing, the emitted
state is mixed. Moreover, even an emitted pure manybody state
generally gives rise to a mixed reduced one- or two particle state. It
is therefore appropriate to discuss the state in terms of density
matrices.

To simplify the discussion we consider a spin-polarized system with
scattering amplitudes independent on energy on the scale of the
applied bias $eV$, i.e. the linear voltage regime. The emitted state
then has only orbital degrees of freedom. We first consider the
single-particle orbital state emitted e.g. towards $A$ (same
considerations hold for $B$). Introducing operators $b_{An}^{\dagger}$
creating electrons in lead $An$, with $n=1,2$, propagating out from
the source, the $2 \times 2$ density matrix (not normalized) is by
definition given by
\begin{equation}
\rho_A=\sum_{n,m=1}^2\rho_{nm} b_{An}^{\dagger}|0\rangle \langle 0|b_{Am}=\left(\begin{array}{cc} \rho_{11} & \rho_{12} \\ \rho_{21} & \rho_{22} \end{array}\right)
\label{onedens}
\end{equation}
where we work in the basis $\{|1\rangle_A,|2\rangle_A\}$, with
$b_{An}^{\dagger}|0\rangle=|n\rangle_A$, formed by the lead indices
(see Fig. \ref{figcond}). The matrix elements $\rho_{nm}=\langle
b_{Am}^{\dagger}b_{An} \rangle$. The Hermitian density matrix,
$\rho_A=\rho_{A}^{\dagger}$, has four independent parameters and can
be written as follows
\begin{equation}
\rho_A=\frac{1}{2}\sum_{i=0}^3 c_i \sigma_i=\frac{1}{2}\left(\begin{array}{cc} c_0+c_3 & c_1-i c_2 \\ c_1+ic_2 & c_0-c_3 \end{array}\right),
\label{paramone}
\end{equation}
where $\{\sigma_i\}=[{\bf 1},\sigma_x,\sigma_y,\sigma_z]$. A
normalized density matrix is obtained by dividing all elements by
$c_0$. In the same way, the two-particle density matrix is given by
\begin{equation}
\rho_{AB}=\sum_{n,m,k,l=1}^2\rho_{nm}^{kl} b_{An}^{\dagger}b_{Bk}^{\dagger}|0\rangle \langle 0|b_{Bl}b_{Am}
\end{equation}
with the matrix elements $\rho_{nm}^{kl}= \langle
b_{Am}^{\dagger}b_{Bl}^{\dagger}b_{Bk}b_{An}\rangle$. The two-particle
density matrix has 16 independent parameters and can be written
\begin{eqnarray}
\rho_{AB}=\frac{1}{4}\sum_{i,j=0}^3c_{ij}\sigma_i\otimes \sigma_j
\label{twopartdens}
\end{eqnarray}
with $\otimes$ the direct product. Expressing the real coefficients
$c_{i}$ and $c_{ij}$ in terms of outcomes of ensemble averaged
measurements thus gives a complete reconstruction of the emitted
state. The accessible measurements are average current and zero
frequency current correlations. Importantly, in the transport state
the source continuously emits quantum states. As a consequence, the
long time measurements automatically provide an ensemble average
measurement.  At $A$ the average currents at contacts $\alpha=\pm$ are
\begin{eqnarray}
I_A^{\alpha}=\frac{e^2V}{h}\langle n_{A}^{\alpha}\rangle, \hspace{0.5cm} n_A^{\alpha}= b_{A\alpha}^{\dagger}b_{A\alpha}.
\label{curr}
\end{eqnarray}
The {\it zero frequency} correlator between currents fluctuations $\Delta I$
in reservoirs $A\alpha$ and $B\beta$, given by
$S_{AB}^{\alpha\beta}=(1/2)\int dt \langle \Delta I_{A\alpha}(0)\Delta
I_{B\beta}(t)+ \Delta I_{B\beta}(t)\Delta I_{A\alpha}(0)\rangle$ can
be written \cite{Buttnoise}
\begin{eqnarray}
S_{AB}^{\alpha\beta}&=&\frac{2e^3V}{h}\left[\langle n_{A}^{\alpha}n_{B}^{\beta}\rangle-\langle n_{A}^{\alpha}\rangle\langle n_{B}^{\beta}\rangle\right]
\label{noise}
\end{eqnarray}
with $n_B^{\alpha}= b_{B\alpha}^{\dagger}b_{B\alpha}$. The operators
$b_{A\alpha}$ and $b_{B\beta}$ in the reservoirs at $A$ and $B$ are
related to operators $b_{An}$ and $b_{Bk}$ in the leads $An,Bk$, with
$n,k=1,2$, (see Fig. \ref{figcond}) via the scattering matrix $S_A$ of
the beamsplitters at $A$ as
\begin{eqnarray}
\left(\begin{array}{c}b_{A+} \\ b_{A-} \end{array}\right)=
S_A\left(\begin{array}{c}b_{A1} \\ b_{A2} \end{array}\right)
\label{smatrel}
\end{eqnarray}
and similarly at $B$. 

We start with the reconstruction of the one-particle state at $A$,
accessible via the average current (the same procedure holds for the
state at $B$). Here a formal approach is taken which directly can be
extended to the investigation of the two-particle state. We note that
the reconstruction approach is similar to QST schemes for qubits in
other systems, see e.g. Refs. \cite{Kwiat2,Nori}. There are however a
number of important special features for mesoscopic systems, making a
detailed investigation important. It is desirable to minimize both the
type and number of experiments having to be carried out. As is clear
from the following, for a complete reconstruction it is sufficient to
consider only measurements of currents in one reservoir in $A$. Here
we consider the current at $A+$. Using the relation between operators,
Eq. (\ref{smatrel}), and Eq. (\ref{curr}), we have
\begin{eqnarray}
I_A^+/(e^2V/h)=\langle n_A^+\rangle=\mbox{tr}\left(\rho_A{\cal A}\right)
\label{currplus}
\end{eqnarray}
with the matrix
\begin{eqnarray}
{\cal A}=\left(\begin{array}{cc} R_A & \sqrt{T_AR_A}e^{-i(\phi_A+\varphi_A)} \\ \sqrt{T_AR_A}e^{i(\phi_A+\varphi_A)}  & T_A  \end{array}\right).
\label{amat}
\end{eqnarray}
The phase $\varphi_A=\phi_{A2}-\phi_{A3}$ contains all the information
about uncontrollable phases of the beamsplitter.  From Eqs.
(\ref{currplus}) and (\ref{amat}) it is clear that the phase
$\varphi_A$ can be included in $\rho_A$ by a change of local basis
$\rho_A\rightarrow U_A \bar\rho_A U_A^{\dagger}$ with
$U_A=\mbox{diag}[\mbox{exp}(-i\varphi_A/2),\mbox{exp}(i\varphi_A/2)]$. Below
we consider the reconstruction of $\bar \rho_A$, parameterized by
coefficients $\bar c_{i}$ [see Eq. (\ref{paramone})], thus working
with ${\mathcal A}(\varphi_A=0)$ in Eq. (\ref{amat}). This yields
$\rho_A$ up to an unknown local basis rotation.

\begin{figure}[b]   
\centerline{\psfig{figure=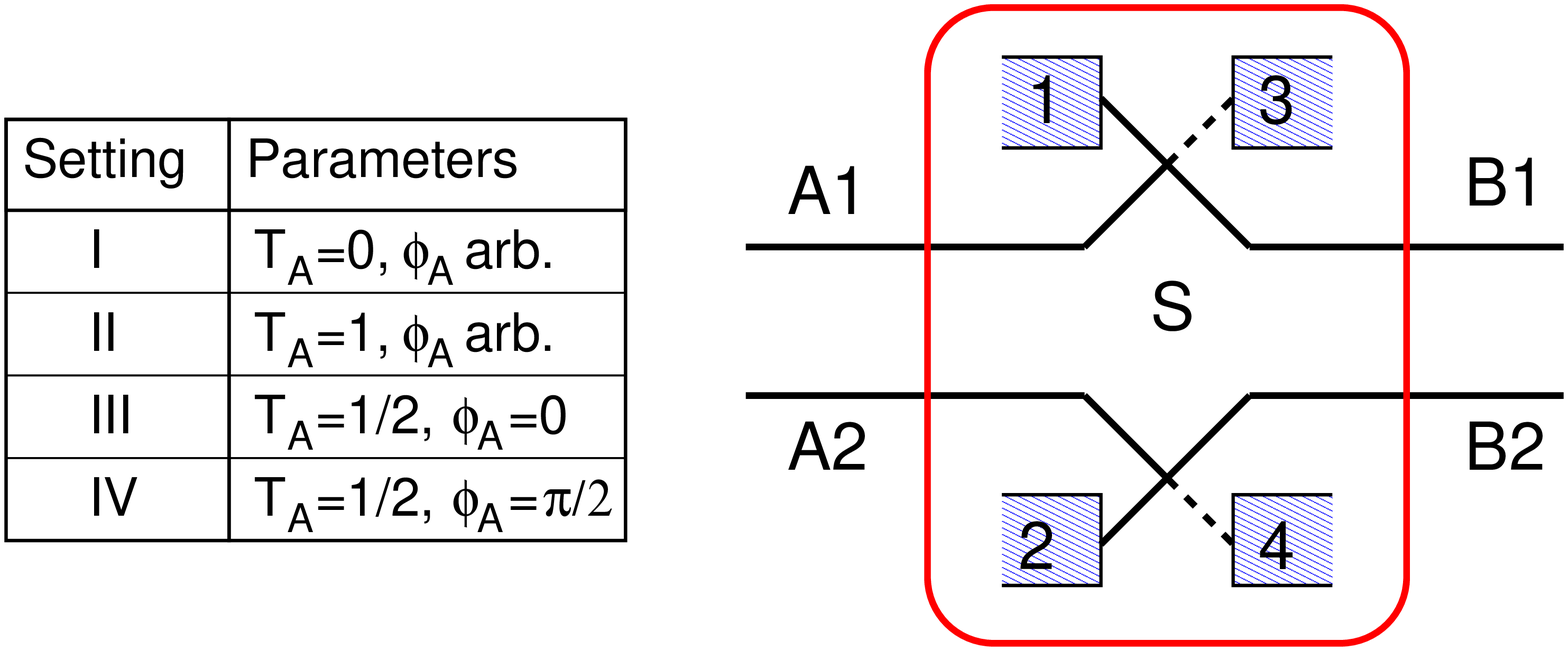,width=8.5cm}}
\caption{Left: Table with scattering parameters at $A$ for the four
different settings. Right: Schematic of an elementary mesoscopic
source S (red box, see Fig. \ref{figcond}). Two reservoirs biased at
$eV$, $1$ and $2$, and two grounded reservoirs, $3$ and $4$ are
connected via beamsplitters to the four leads going out towards $A$
and $B$. Scattering between upper and lower leads, e.g. $A1$ and $A2$,
is not possible.}
\label{fig2}
\end{figure}

Importantly, only four settings of the beamsplitter are needed, both
for the current and the current correlators, for a complete state
reconstruction. The settings $I$ to $IV$ considered here are listed in
the table in Fig. \ref{fig2}.
By constructing suitable linear combinations $j_A(j)$ of the
observables at the different settings, in the
$\{|1\rangle_A,|2\rangle_A\}$ basis
\begin{eqnarray}
j_A(0)&=&{\mathcal A}(I)+{\mathcal A}(II)={\bf 1}, \nonumber \\ 
j_A(1)&=&2{\mathcal A}(III)-\left[{\mathcal A}(I)+{\mathcal A}(II)\right]=\sigma_x, \nonumber \\
j_A(2)&=&2{\mathcal A}(IV)-\left[{\mathcal A}(I)+{\mathcal A}(II)\right]=\sigma_y, \nonumber \\
j_A(3)&=&{\mathcal A}(I)-{\mathcal A}(II)=\sigma_z, 
\label{orthobs}
\end{eqnarray}
we obtain a complete set \cite{Fano} of measurements, since the Pauli
matrices $\sigma_j$ obey the relation
$\mbox{tr}(\sigma_i\sigma_j)=2\delta_{ij}$. Here ${\mathcal A(I)}$ is
the matrix ${\mathcal A}$ in Eq. (\ref{amat}) for the setting I
etc. From Eqs. (\ref{currplus}) and (\ref{orthobs}) and the relation
$\langle j_A(j)\rangle=\mbox{tr}(\bar \rho_A \sigma_j\rangle=\bar
c_j$ we then directly obtain the coefficients $\bar c_j$,
parametrizing $\rho_A$ in Eq. (\ref{paramone})
\begin{eqnarray}
\bar c_j&=&\sum_{k=0}^3Q_{jk}\langle n_A^+(k)\rangle, \hspace{0.1 cm} Q=\left(\begin{array}{cccc} 1 & 1&0&0 \\ 
-1  & -1  & 2 &0 \\ -1 & -1 & 0 &2 \\ 1 &-1&0&0 \end{array}\right)
\label{crel}
\end{eqnarray}
in terms of the measured currents for the different settings, taking
the index $\{k\}=[0,1,2,3]\equiv[I,II,II,IV]$. This completes the
one-particle state reconstruction.
   
We then turn to the two-particle state. In Eq. (\ref{noise}), the
quantity that is directly linked to the density matrix elements
$\rho_{nm}^{kl}$ is the reducible correlator $\langle
n_{A}^{\alpha}n_{B}^{\beta}\rangle$. This correlator is directly
obtained from the measured noise and the average currents. In analogy
to the current, it is sufficient to consider correlations between
currents in one terminal in $A$ and one in $B$. Considering here $A+$
and $B+$, one obtains from Eq. (\ref{noise}) and (\ref{smatrel}) the
dimensionless correlator
\begin{eqnarray}
\frac{S_{AB}^{++}}{2e^3V/h}+\frac{I_{A}^+I_B^+}{(e^2V/h)^2}=\langle n_{A}^{+}n_{B}^{+}\rangle=\mbox{tr}\left(\rho_{AB}{\cal A}\otimes{\cal B}\right)
\label{noiseplus}
\end{eqnarray}
where the matrix ${\cal B}$ is given from ${\cal A}$ in
Eq. (\ref{amat}) by changing indices $A \rightarrow B$ in the
scattering amplitudes. Similar to the one-particle state, we note from
Eq. (\ref{noiseplus}) that both phases $\varphi_A$ and $\varphi_B$ can
be included in $\rho_{AB}$ by independent local rotations $\rho_{AB}
\rightarrow (U_A\otimes U_B)\bar\rho_{AB} (U_A\otimes U_B)^{\dagger}$,
with
$U_B=\mbox{diag}[\mbox{exp}(-i\varphi_B/2),\mbox{exp}(i\varphi_B/2)]$.
Below we thus consider the reconstruction of $\bar\rho_{AB}$,
parameterized as in Eq. (\ref{twopartdens}) by the coefficients $\bar
c_{ij}$, yielding $\rho_{AB}$ up to a local basis
rotation \cite{entcom}.

By considering the same four settings at $B$ as at $A$, we can use the
linear combination operators $j_A(j)$ in Eq. (\ref{orthobs}) and
correspondingly $j_B(i)$ to construct a complete set of observables,
in the basis
$\{|1\rangle_A|1\rangle_B,|1\rangle_A|2\rangle_B,|2\rangle_A|1\rangle_B,|2\rangle_A|2\rangle_B\}$,
\begin{eqnarray}
j_A(j)j_B(i)=\sigma_{j}\otimes\sigma_{i}
\end{eqnarray}
since the direct products of $\sigma$-matrices obey
$\mbox{tr}\left[(\sigma_j\otimes
\sigma_i)(\sigma_k\otimes\sigma_l)\right]=4\delta_{jk}\delta_{il}$. From
Eq. (\ref{noiseplus}) and the relation $\langle
j_A(j)j_B(i)\rangle=\mbox{tr}(\bar \rho_{AB}
\sigma_j\otimes\sigma_i\rangle=\bar c_{ji}$ we then directly obtain
the coefficients $\bar c_{ji}$ as
\begin{eqnarray}
\bar c_{ji}=\sum_{k,l=0}^3Q_{jk}Q_{il}\langle n_A^+(k)n_B^+(l)\rangle
\label{twocoeff}
\end{eqnarray}
in terms of the measured current correlators and averaged currents. We
emphasize that all elements can be determined from sixteen current
correlations and eight average currents (four at $A$ and four at $B$).
We also note that the reconstructed density matrix, due to nonideal
measurements, might have negative eigenvalues, i.e. it might not be
positive semidefinite. Schemes to correct for this for one and
two-qubit states are discussed in e.g. ref. \cite{Kwiat2}.

In the context of two-particle entanglement, it is interesting to
compare the QST-scheme with a Bell Inequality, recently discussed for
mesoscopic system (see e.g. refs. \cite{Cht,Orb} and for a density
matrix approach ref. \cite{Carlodeph}). Both schemes require the same
number of current correlation measurements. The density matrix
reconstructed by QST however completely determines the
entanglement. In contrast, a Bell Inequality can not be used to
quantify the entanglement \cite {Verstraete}, there are e.g. mixed
entangled states \cite{Werner} that do not lead to a violation of a
Bell Inequality.

It is clarifying to illustrate the above scheme with a simple example
(see Fig. \ref{fig2}). We consider the Hanbury Brown Twiss geometry of
ref. \cite{QH2}, where the number of nonzero elements of the one and
two-particle density matrices are reduced due to the topological
properties of the conductor. Since no scattering between the upper,
$1$, and lower, $2$, leads is physically possible due to the spatial
separation, the one-particle density matrix $\bar \rho_A$ only has two
nonzero elements, $\bar \rho_{11}$ and $\bar \rho_{22}$. These
elements are parametrized by $\bar c_0$ and $\bar c_3$, obtained by
measuring $\langle j_A(0)\rangle$ and $\langle j_A(3)\rangle$.

The two-particle density matrix $\bar \rho_{AB}$ has four nonzero
elements, $\bar \rho_{11}^{22},\bar \rho_{22}^{11},\bar
\rho_{21}^{12}$ and $\bar \rho^{21}_{12}$. Using the relation between
the coefficients $\bar c_{ij}$ resulting from several matrix elements
being zero, $\bar \rho_{AB}$ can then be parameterized as
\begin{equation}
\bar \rho_{AB}=\frac{1}{2}\left(\begin{array}{cccc} 0 &0&0&0 \\ 0 &
\bar c_{00}+\bar c_{33} & \bar c_{11}-i\bar c_{21} & 0 \\ 0& \bar
c_{11}+i\bar c_{21} & \bar c_{00}-\bar c_{33} &0 \\ 0 &0&0&0 \end{array}\right).
\end{equation}
Conseqently, only four correlations $\langle j_A(i)j_B(j)\rangle$ need
to be measured to completely reconstruct $\bar \rho_{AB}$, reducing
the number of actual current correlations needed to be carried out to
12 for the settings considered here [see Eq. (\ref{twocoeff})]. It is
interesting to note that in the geometry in Fig. \ref{fig2},
considering the tunneling limit for the beamsplitters in the source
and changing to an electron-hole picture \cite{QH1}, the emitted state
is a true two-particle state \cite{QH2}. Since the hole currents and
current fluctuations are directly related to the electron ones, it is
possible to employ our scheme to reconstruct an electron-hole state as
well.

In conclusion, we have presented a scheme for orbital Quantum State
Tomography, a complete reconstruction of orbital quantum states in
mesoscopic conductors in the transport state. The emitted orbital
states are manipulated with electronic beamsplitters and detected with
currents and zero frequency current correlations. With all components
experimentally available, our scheme opens up for a direct observation
of quasiparticle quantum states in mesoscopic conductors.

We acknowledge discussions with E.V. Sukhorukov. This work was
supported by the Swedish Research Council, the Swiss National Science
Foundation and the program for Materials with Novel Electronic
Properties.


\begin{thebibliography}{99}
\bibitem{Pauli}
Historically, W. Pauli, {\it General 
Principles of Quantum Mechanics} (Springer, Berlin, 1980) addressed the problem for position and momentum measurements in 1933. 
\bibitem{Fano}
U. Fano, Rev. Mod. Phys. {\bf 29}, 74 (1957).
\bibitem{QSTrevs}
J. Bertrand and P. Bertrand, Found. Phys. {\bf 17}, 397 (1987); K. Vogel and H. Risken, Phys. Rev. A {\bf 40}, 2847 (1989); A. Royer, Found. Phys. {\bf 19}, 3 (1989); U. Leonhardt, Phys. Rev. B {\bf 53}, 2998 (1996); M. G. Raymer, Cont. Phys. {\bf 38}, 343 (1997).
\bibitem{Ashburn}
J.R. Ashburn {\it et al}, Phys. Rev. A {\bf 41}, 2407 (1990).
\bibitem{Smithey}
D.T. Smithey {\it et al}, Phys. Rev. Lett. {\bf 70}, 1244 (1993).
\bibitem{Dunn}
T.J. Dunn, I.A. Walmsey and S. Mukamel, Phys. Rev. Lett. {\bf 74}, 884 (1995).
\bibitem{Liebfried}
D. Liebfried {\it et al},  Phys. Rev. Lett. {\bf 77}, 4281 (1996).
\bibitem{Kurtsiefer}
Ch. Kurtsiefer, T. Pfau, and Mlynek, Nature {\bf 386}, 150 (1997).
\bibitem{EntQST}
P.G. Kwiat {\it et al}, Nature {\bf 409} 1014 (2001); T. Yamamoto {\it et al}, 
{\it ibid} {\bf 421}, 343 (2003).
\bibitem{Nori}
Y. Liu, L.F. Wei, and F. Nori, Europhys. Lett. {\bf 67}, 187 (2004); quant-ph/0407197.
\bibitem{Orb}
P. Samuelsson, E.V. Sukhorukov, and M. B\"uttiker,
Phys. Rev. Lett. {\bf 91}, 157002 (2003).
\bibitem{QH1} 
C.W.J. Beenakker {\it et al}, Phys. Rev. Lett. {\bf 91}, 147901 (2003).
\bibitem{QH2}
P. Samuelsson, E.V. Sukhorukov, and M. B\"uttiker, Phys. Rev. Lett. {\bf 92}, 026805 (2004).
\bibitem{BS1} 
M. Henny {\it et al.}, Science {\bf 284}, 296 (1999); S. Oberholzer
{\it et al.}, Physica {\bf 6E}, 314 (2000). 
\bibitem{BS2}
W.D. Oliver {\it et al.},  Science {\bf 284}, 299 (1999). 
\bibitem{Buttnoise}
M. B\"uttiker, Phys. Rev. B {\bf 46}, 12485 (1992).
\bibitem{Buttrev}
Ya. Blanter and M. B\"uttiker, Phys. Rep. {\bf 336}, 1 (2000).
\bibitem{QHexp}
Y. Ji {\it et al.}, Nature {\bf 422,} 415 (2003).
\bibitem{Sukh}
G. Burkhard, D. Loss, and E.V. Sukhorukov, 
Phys. Rev. B {\bf 61}, 16303 (2000). 
\bibitem{Cht}
N. Chtchelkatchev {\it et al}, Phys. Rev. B {\bf 66}, 161320 (2002).  
\bibitem{SSQI}
See e.g. P. Samuelsson, E.V. Sukhorukov, and M. B\"uttiker, cond-mat/0503016 for a recent discussion.
\bibitem{Kwiat2} D. James {\it et al}, Phys. Rev. A {\bf 64}, 052312
(2001).
\bibitem{entcom}
Note that e.g. the entanglement is invariant under such a local rotation $\rho_{AB}
\rightarrow (U_A\otimes U_B)\bar\rho_{AB} (U_A\otimes U_B)^{\dagger}$.
\bibitem{Carlodeph} 
J.L. van Velsen {\it et al}, Turk. J. Phys. {\bf
27}, 323 (2003).
\bibitem{Verstraete}
F. Verstraete and M. Wolf, Phys. Rev. Lett. {\bf 89}, 170401 (2002).
\bibitem{Werner}
R.F. Werner, Phys. Rev. A {\bf 40}, 4277 (1989).
\end{thebibliography}
\end{document}